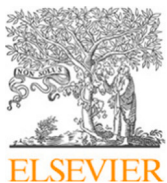
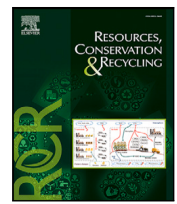

Full Length Article

# Using hyperspectral imaging to identify optimal narrowband filter parameters for construction and demolition waste classification

Stanislav Vítek [a], Tomáš Zbíral [b], Václav Nežerka [b,*]

[a] *Faculty of Electrical Engineering, Czech Technical University in Prague, Technická 2, 166 27 Praha 6, Czech Republic*
[b] *Faculty of Civil Engineering, Czech Technical University in Prague, Thákurova 7, 166 29 Praha 6, Czech Republic*

A B S T R A C T

Hyperspectral imaging (HSI) is widely applied in various industries, enabling detailed analysis of material properties or composition through their spectral signatures. However, for classification of construction and demolition waste (CDW) materials, HSI is impractical since real-time sorting requires rapid data acquisition and lightweight classification. Instead, fitting selected narrowband filters onto standard cameras can achieve comparable results with substantially reduced computational overhead. In this study, reflectance data of common CDW materials were recorded using a hyperspectral camera, and a multilayer perceptron classifier was employed to evaluate different feature sets. The findings indicate that adding only two wavelengths beyond the RGB channels is sufficient for high-accuracy classification, with optimal filter central wavelengths identified at approximately 650–750 nm and 850–1000 nm across the tested bandwidths (5–50 nm) highlighting the importance of near-infrared regions for material discrimination.

## 1. Introduction

Managing construction and demolition waste (CDW) is a significant global challenge with substantial environmental and economic implications. CDW accounts for a large proportion of the total waste generated worldwide, and improper handling and disposal can lead to environmental degradation and resource depletion. The construction industry, which produces around 25% of the global GDP and employs 7% of the population (Norouzi et al., 2021), is responsible for the enormous consumption of raw materials and the large-scale production of waste. Gravel, sand, and crushed stone represent the most extracted category (Valentini, 2023), highlighting the construction sector's dependence on these key raw materials. It is estimated that the construction sector consumes over 30%–40% of all natural resources extracted globally (Darko et al., 2020; Purchase et al., 2021), produces around 25%–40% of the total solid waste (Nasir et al., 2017), and contributes up to 25% of anthropogenic $CO_2$ emissions (Mahpour, 2018). In the European Union (EU) alone, approximately 18 million people were employed in the construction sector in 2020, and the production of CDW was estimated to be around 747.3 million tons, equivalent to approximately 1 685 kg per capita.[1] Effective sorting and recycling of CDW are crucial to minimizing waste, conserving natural resources, and promoting sustainable construction practices (İlcan et al., 2024b,a). However, the classification of different materials present in CDW streams remains a complex task, often requiring labor-intensive and error-prone manual sorting methods.

Current methods for classifying CDW materials, such as manual sorting and conventional imaging techniques, face several limitations. Manual sorting is costly, time-consuming, and prone to human error. Traditional imaging techniques, e.g., color-based classification (Sulistiyowati et al., 2024), can help identify some materials but often struggle to differentiate visually similar materials. These limitations underscore the need for more advanced, automated approaches that can accurately classify a wide range of materials found in CDW. As highlighted by Hoong et al. (2020), improper sorting is a major limiting factor in the valorization of crushed CDW, particularly when using recycled aggregates in new concrete mixes or as micro-fillers. Research into the development of more sophisticated classification methods, such as those based on robotic vision and machine learning, therefore holds significant potential to enhance CDW recycling and reuse (Su, 2020; Davis et al., 2021).

Advanced methods have been developed to sort CDW using computer vision and robotics. Chen et al. (2022) introduced a robotic system for automatic waste sorting, utilizing deep learning and simultaneous localization and mapping technology. Their approach focuses on real-time navigation and object grasping in complex environments using RGB-depth cameras and 3D LiDAR, emphasizing physical waste

---

handling rather than spectral data analysis. Similarly, Dong et al. (2022) proposed a boundary-aware transformer model for fine-grained recognition of mixed waste materials, enhancing texture-based approaches with crucial set features and expanding the potential for use in robotics applications. Their study also relies on image-based methods, contrasting with our focus on spectral reflectance data.

There has also been a growing interest in leveraging hyperspectral imaging (HSI) for material classification in waste management (Xiao et al., 2019a; Ku et al., 2020; Lan et al., 2024). HSI offers a promising solution due to its ability to capture detailed spectral information across various wavelengths, providing unique spectral signatures for different materials. Unlike traditional imaging, which captures data in only three broad bands (red, green, and blue), hyperspectral cameras collect data across hundreds of narrow, contiguous wavelength bands, allowing for the differentiation of materials with similar visual appearances but distinct spectral characteristics (Lu and Chen, 2022). Previous studies have demonstrated the effectiveness of HSI in various fields, including material science (Florkowski, 2020; Ichi and Dorafshan, 2023), agriculture (Marston et al., 2022; Nguyen et al., 2024), food industry (Noviyanto and Abdulla, 2019; Han and Gao, 2019), medicine (Petracchi et al., 2024), and also solid waste management (Serranti and Bonifazi, 2010), where it has been used mostly to distinguish between different types of plastics (Pieszczek and Daszykowski, 2019; Zheng et al., 2018; Taneepanichskul et al., 2023).

Even though full-spectrum hyperspectral analysis can yield high classification accuracy, it remains impractical for real-time applications due to lengthy data acquisition and substantial computational overhead. Instead, employing a minimal set of strategically selected narrowband filters on standard monochrome cameras offers a more cost-effective solution. While this approach has not yet been applied to CDW classification, it has proven effective in other domains, such as precision agriculture with multispectral cameras mounted on UAVs (Deng et al., 2018).

In the context of CDW management, the closest study to our work is by Xiao et al. (2019b,a), who achieved high accuracy using hyperspectral imaging and machine learning. However, their research focused on relatively simple materials, distinguishable with conventional methods, and relied on the entire spectral range for feature extraction. In contrast, our study aims to identify only the minimal number of key wavelengths necessary to capture critical differences among more challenging CDW materials, facilitating cost-effective classification systems that operate with fewer filters.

This study introduces a novel two-stage approach designed to balance classification accuracy with practical feasibility in industrial environments. In the first stage, it is demonstrated that adding only two wavelengths beyond the RGB channels substantially enhances classification performance, thus removing the necessity for full-spectrum data. Building upon these results, the second stage systematically evaluates narrowband filters at central wavelengths spanning 425–975 nm in 25 nm increments and with full-width at half-maximum (FWHM) bandwidths of 5, 20, 35, and 50 nm. By examining these parameters, the proposed approach provides actionable guidelines for selecting commercially available or custom-manufactured filters, ensuring that insights derived from hyperspectral imaging translate directly into large-scale, real-time sorting lines. As a result, this contribution represents a significant advance toward integrating hyperspectral-derived intelligence into industrial CDW management practices.

## 2. Materials and methods

### 2.1. Materials

The materials selected for this study (presented in Fig. A.7, Appendix section Appendix A) were chosen in collaboration with the managers of ENVISAN-GEM, a Czech company specializing in the collection, recycling, and transport of construction debris, as well as waste disposal and earthworks. The selected materials are commonly found in pre-sorted CDW collected from demolition or reconstruction sites. These materials present significant challenges for accurate sorting, as many exhibit visual similarities in color, texture, and shape, making them difficult to differentiate, even for human operators on sorting lines or advanced machine learning models based on RGB imaging (Nežerka et al., 2024). Fig. 1 illustrates the large quantities of these pre-sorted and crushed materials at the sorting yard. Despite initial sorting efforts, these CDW heaps contain incompatible materials with varying chemical compositions, durability, and strength. This heterogeneity limits the full utilization of the recycled materials, often relegating them to low-value applications such as fillings or backfills. By effectively sorting these mixed materials using the methods investigated in this study, it becomes possible to enhance the quality and value of the recycled products, enabling their use in higher-grade applications like aggregates for new concrete production (Ying et al., 2022; Russo and Lollini, 2022).

Critical materials include aerated-autoclaved concrete (AAC), standard concrete, and mortar, which are visually similar but have distinct recycling values. Concrete contains valuable aggregates that are easily accessible after crushing and can serve as substitutes for primary materials. In contrast, the inclusion of AAC as a recycled aggregate would degrade the properties of the resulting composites (Suwan and Wattanachai, 2017; He et al., 2020) or a special treatment of such aggregates would be needed (Ji et al., 2024). Despite their similar chemical compositions, clay bricks and ceramic roof tiles also pose challenges in sorting. Clay bricks exhibit lower strength than ceramic roof tiles. Yet, they can be easily confused due to their similar appearances, despite differing structural properties and eventually performance when used as recycled aggregates (Miličević et al., 2016).

Including lightweight materials, such as extruded polystyrene (EPS) and wood, presented an additional challenge. Although these materials could theoretically be distinguished by gravimetric methods (e.g., EPS could be separated using blowers), integrating them into a single sorting line with other materials would be more practical for industrial applications or on-site autonomous robotic sorting systems.

Glazed ceramic tiles, frequently used as bathroom cladding, were another critical material studied. Although strong, their smooth surface hinders proper bonding when used as recycled aggregates. Since only one side of the tiles is glazed, both sides needed to be imaged for appropriate classification. These tiles were also selected for their variability in color and texture, making them the most challenging material to classify using optical sensors. We anticipated that the accuracy of classification for this material would be relatively lower.

### 2.2. Hyperspectral imaging

The reflectance of the CDW fragments was measured using the SPECIM PFD4K-65-V10E hyperspectral camera (Fig. B.8, Appendix section Appendix B) covering wavelengths from 400 to 1 000 nm. Although the camera's spectral resolution is approximately 3 nm, it provides 768 distinct spectral bands across this range, resulting in a spectral sampling interval of about 0.78 nm. This fine spectral sampling facilitates detailed spectral analysis, even though the instrument's ability to resolve closely spaced wavelengths is limited by its spectral resolution. The camera is equipped with OLE 23 optics, featuring a focal length of 23 mm and an f-number of f/2.4, ensuring high spatial resolution with an RMS spot size of less than 9 μm.

Each CDW sample was illuminated using a halogen lamp to provide consistent and uniform illumination. The hyperspectral camera captured the reflected light from the sample, generating a raw hyperspectral image data cube, $I_{HS}(x, y, \lambda)$, where $x$ and $y$ are the spatial coordinates of the pixels, and $\lambda$ is the spectral coordinate representing the wavelength (Zahra et al., 2024).

The key objective was to obtain the normalized reflectance, $I_{norm}(\lambda)$, which represents the ratio of the reflected light to the incident light at each wavelength $\lambda$. The raw data required radiometric calibration to ensure accurate reflectance measurements; the process was in detail described by Krauz et al. (2022). We acquired dark and white reference images as part of this calibration process. Dark images, represented by $I_{\text{DARK}}(x, y, \lambda)$, were captured with the camera shutter closed, which accounted for the sensor's dark current noise. White reference images, $I_{\text{WHITE}}(x, y, \lambda)$, were obtained using a calibration target with a reflectance close to 100%, providing a baseline for normalization.

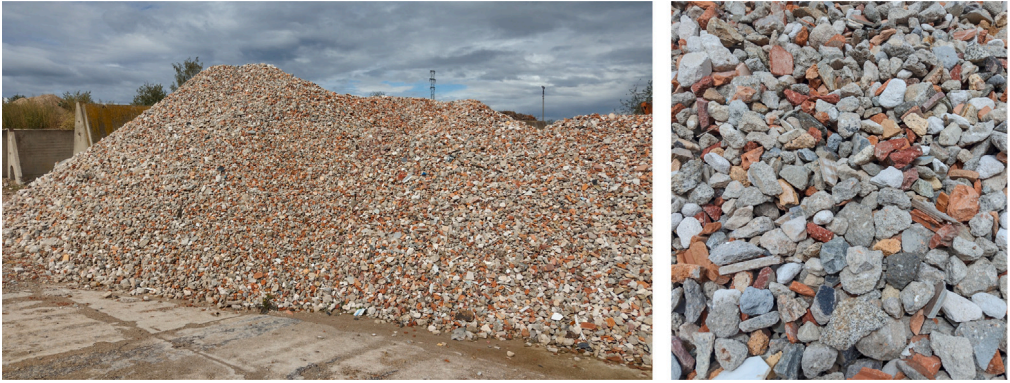

**Fig. 1.** Large quantities of pre-sorted and crushed CDW materials at the Envisan GEM sorting yard in South Bohemia, Czech Republic. The heaps contain mixed materials that require further sorting to improve their utilization potential and enable higher-value recycling applications.

The normalized reflectance, $I_{\text{norm}}(x, y, \lambda)$, was computed by first subtracting the dark image from the raw hyperspectral image and then dividing by the difference between the white reference image and the dark image:

$$I_{\text{norm}}(x, y, \lambda) = \frac{I_{\text{HS}}(x, y, \lambda) - I_{\text{DARK}}(x, y, \lambda)}{I_{\text{WHITE}}(x, y, \lambda) - I_{\text{DARK}}(x, y, \lambda)}. \quad (1)$$

This normalization corrects for any variations in sensor response and illumination intensity, resulting in reflectance values independent of these factors.

### 2.3. Feature extraction

To achieve accurate classification of CDW materials in a manner suitable for industrial application, it is necessary to extract features from the hyperspectral measurements that retain each material's distinctive spectral characteristics, yet do so without introducing excessive complexity or computational costs. This feature extraction approach is implemented in two stages to ensure both effective discrimination and practical feasibility.

In Stage 1, the focus was on minimizing the number of additional wavelengths beyond the standard RGB channels, thereby maintaining high classification accuracy without resorting to a full-spectrum analysis. In Stage 2, a systematic evaluation of narrowband filters with different central wavelengths and bandwidths was conducted.

#### 2.3.1. Stage 1: Determination of essential wavelengths

In Stage 1, the primary objective was to determine the minimal number of discrete wavelengths required to significantly improve classification accuracy beyond the standard RGB channels. To achieve this, mean reflectance curves were derived for each material subset by averaging the normalized reflectance values across all pixels at each wavelength $\lambda$:

$$I_{\text{norm}}(\lambda) = \frac{1}{N} \sum_{(x,y) \in \text{subset}} I_{\text{norm}}(x, y, \lambda), \quad (2)$$

where $N$ is the total number of pixels in the subset, and $I_{\text{norm}}(x, y, \lambda)$ is the normalized reflectance at pixel $(x, y)$ for wavelength $\lambda$. The resulting mean reflectance curve $I_{\text{norm}}(\lambda)$ represents the average spectral behavior of the material.

We selected a set of wavelengths that included the standard red, green, and blue (RGB) spectral bands and additional wavelengths at regular intervals across the spectrum. Specifically, the selected wavelengths were:

$$\Lambda = \{400, 415, 500, 540, 600, 660, 700, 800, 900, 1000\} \text{ nm}. \quad (3)$$

These wavelengths provide coverage of the spectral range from 400 nm to 1000 nm, allowing us to capture key features of the materials' reflectance properties within this measured range.

From these selected wavelengths, we extracted features to use in our classification models as described next.

*Reflectance values at RGB wavelengths.* We began by focusing on the wavelengths corresponding to the blue, green, and red spectral bands, identified as $\lambda_B = 415$ nm (blue), $\lambda_G = 540$ nm (green), and $\lambda_R = 660$ nm (red) (Wang et al., 2017). For each of these wavelengths, we extracted the mean normalized reflectance values:

$$F_B = I_{\text{norm}}(\lambda_B), \quad F_G = I_{\text{norm}}(\lambda_G), \quad F_R = I_{\text{norm}}(\lambda_R). \quad (4)$$

These features represent how the materials reflect light in the primary color bands and are readily accessible using standard RGB imaging equipment.

*Inclusion of additional wavelengths.* We incrementally added reflectance values at additional wavelengths to capture spectral characteristics not apparent in the RGB bands. The selected additional wavelengths were:

$$\Lambda' = \{400, 500, 600, 700, 800, 900, 1000\} \text{ nm}. \quad (5)$$

For each wavelength $\lambda$ in $\Lambda'$, we defined a single-wavelength feature:

$$F_\lambda = I_{\text{norm}}(\lambda). \quad (6)$$

We started by adding one additional wavelength from $\Lambda'$ to the RGB feature set and progressively included more wavelengths. This approach allowed us to analyze how the number and selection of wavelengths influence the accuracy of the machine learning models. Feature sets were created by combining the RGB wavelengths with additional wavelengths from $\Lambda'$, for example:

- Feature set with RGB and one additional wavelength: $\{F_B, F_G, F_R, F_{\lambda_1}\}$, where $\lambda_1 \in \Lambda'$.
- Feature set with RGB and all additional wavelengths: $\{F_\lambda \mid \lambda \in \Lambda\}$.

By systematically increasing the number of wavelengths in the feature set, we aimed to determine the impact of additional spectral information on classification performance. Considering the seven additional wavelengths, the total number of possible combinations of these wavelengths (excluding the case where no additional wavelength is selected)

is given by:

$$\sum_{k=1}^{7} \binom{7}{k} = 2^7 - 1 = 127. \tag{7}$$

*Global spectral features.* Finally, most comprehensive models were also trained using features that utilize the entire spectral range:

- Wavelength at peak reflectance:

$$F_{\text{peak}} = \lambda^* \quad \text{where} \quad I_{\text{norm}}(\lambda^*) = \max_{\lambda} I_{\text{norm}}(\lambda). \tag{8}$$

- Area under the reflectance curve:

$$F_{\text{area}} = \int_{400 \text{ nm}}^{1000 \text{ nm}} I_{\text{norm}}(\lambda) \, d\lambda. \tag{9}$$

*2.3.2. Stage 2: Design of optimal narrowband filters*

In Stage 2, the research was built upon the findings of Stage 1 by exploring the use of narrowband filters with various central wavelengths $\lambda_c$ and bandwidths. The goal was to translate the results into practical guidelines for selecting specific filters in industrial sorting application by investing narrowband filters with $\lambda_c$ ranging from 425 nm to 975 nm, in increments of 25 nm. For each $\lambda_c$, filters with FWHM ranging from 5 nm to 50 nm, in increments of 15 nm. This systematic approach allowed to evaluate a wide range of filter configurations:

$$\lambda_c = 425 \text{ nm}, 450 \text{ nm}, \ldots, 975 \text{ nm}; \quad \text{FWHM} = 5 \text{ nm}, 20 \text{ nm}, \ldots, 50 \text{ nm}. \tag{10}$$

For each filter configuration, the mean reflectance within the bandwidth was calculated, assuming constant permeability of the filters to light:

$$F_{\lambda_c, \text{FWHM}} = \frac{1}{\Delta \lambda} \int_{\lambda_c - \Delta \lambda / 2}^{\lambda_c + \Delta \lambda / 2} I_{\text{norm}}(\lambda) \, d\lambda, \tag{11}$$

where $\Delta \lambda$ is the FWHM bandwidth of the filter.

New feature sets were created by combining the RGB reflectance values with the averaged reflectance values from the narrowband filters:

$$\{F_B, F_G, F_R, F_{\lambda_{c1}, \text{FWHM}}, F_{\lambda_{c2}, \text{FWHM}}\}, \tag{12}$$

where $\lambda_{c1}$ and $\lambda_{c2}$ are the central wavelengths of the selected filters, and FWHM is their bandwidth.

*2.4. Machine learning*

In both Stage 1 and Stage 2, machine learning models were employed to classify CDW materials based on extracted spectral features. Detailed descriptions of various machine learning methodologies are available in standard references (Géron, 2022; Murphy, 2022). All experiments were carried out on a laptop computer equipped with an AMD Ryzen 7 5800H CPU with 3.20 GHz (4 cores), 16 GB RAM, a 250 GB SSD, running a 64-bit Windows 11 system and Python 3.10.9. The datasets (the hyperspectral data cubes $I_{\text{norm}}(x, y, \lambda)$), spreadsheets with extracted features, complete results, and Python scripts, including pre-trained models, are provided in a public repository.[2]

The datasets for both Stage 1 and Stage 2 modeling were initially partitioned into training (80%) and testing (20%), with an additional validation split (25% of the training data) used to monitor the loss function during training. Stratification ensured effective model setting and minimized bias in assessing performance on unseen data.

---

[2] https://zenodo.org/records/13840470

*2.4.1. Model architecture and hyperparameters*

A multilayer perceptron (MLP) classifier was selected as the neural network architecture for the classification task. The MLP represents a straightforward yet robust model that can strike a balance between classification accuracy and computational demands, which is critical for practical industrial scenarios. Although other model architectures could potentially yield better performance, the aim of this study is to identify optimal spectral features rather than exhaustively comparing model architectures. The chosen MLP avoids diverting focus from the core objective of selecting narrowband filters that enhance classification efficiency.

The model setting and architecture were guided by previous investigations (Nežerka et al., 2024) and preliminary testing. Initially, an MLP with two hidden layers of 20 neurons each was employed, using the hyperbolic tangent activation function. The weights and biases were optimized via backpropagation with the Adam algorithm (Kingma and Ba, 2015). More complex architectures were also tested, including up to three hidden layers of 50 neurons each and different activation functions. However, these more intricate networks frequently diverged during training unless smaller learning rates were used, which significantly increased training time. Since the objective is to determine the minimal number of wavelengths needed rather than to maximize model performance at all costs, the simpler MLP architecture was retained. This decision did not alter the primary conclusion that two additional discrete wavelengths beyond the RGB channels suffice for effective classification.

The MLP classifier was implemented using the Scikit-Learn v1.1.3 Python library. The training included standardizing the features with the `preprocessing.StandardScaler` class and applying cross-validation via `model_selection.StratifiedShuffleSplit` to ensure stratified sampling across classes. The hyperparameters and model architecture were determined through a grid search optimization process, as summarized in Table D.2 (Appendix section Appendix D). Identical hyperparameters were applied in both Stage 1 and Stage 2 to maintain consistency in comparative analyses.

*2.5. Loss function*

The cross-entropy loss function was used to quantify the discrepancy between predicted class probabilities and actual class labels. For a multi-class classification problem with $C$ classes, the cross-entropy loss $L$ for a single sample $i$ is defined as

$$L_i = -\sum_{k=1}^{C} y_{i,k} \log(p_{i,k}) \tag{13}$$

where $y_{i,k} = 1$ if the true class of sample $i$ is $k$ and 0 otherwise, and $p_{i,k}$ is the predicted probability that sample $i$ belongs to class $k$. The average loss over all $N$ samples is

$$L = \frac{1}{N} \sum_{i=1}^{N} L_i = -\frac{1}{N} \sum_{i=1}^{N} \sum_{k=1}^{C} y_{i,k} \log(p_{i,k}). \tag{14}$$

The training loss $L_{\text{train}}$ and validation loss $L_{\text{val}}$ were recorded at each iteration. The validation loss was computed using the cross-entropy loss on the validation dataset, and the model corresponding to the iteration where $L_{\text{val}}$ attained its minimum was selected.

*2.5.1. Performance evaluation*

The primary evaluation metrics were accuracy ($\alpha$) and weighted F-score ($F_{\text{weighted}}$), selected for their capacity to address any class imbalances.

Accuracy is defined as the proportion of correct predictions among all predictions:

$$\alpha = \frac{\sum_{c=1}^{C} \text{TP}_c}{\sum_{c=1}^{C} \left( \text{TP}_c + \text{FP}_c + \text{FN}_c \right)}, \tag{15}$$

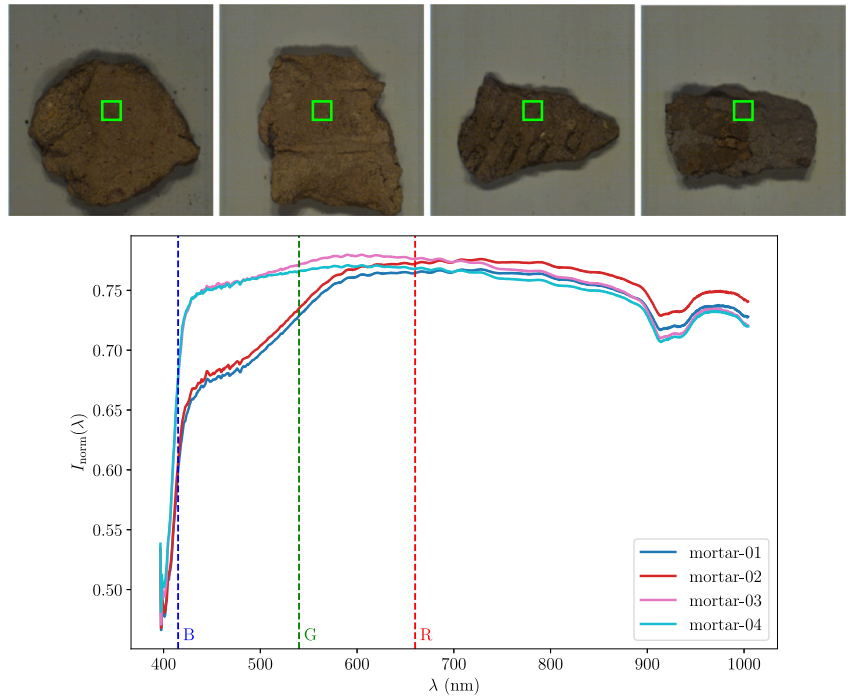

**Fig. 2.** The location of 20 × 20 px subsets used for the extraction of features on mortar samples (top) and the extracted reflectance diagrams with the indication of RGB wavelengths (bottom); the RGB images were reconstructed from the reflectance data at respective wavelengths for the visible light channels.

where $TP_c$ is the number of true positives for class $c$, and $C$ is the total number of classes. Precision ($P_c$) and recall ($R_c$) for each class are

$$P_c = \frac{TP_c}{TP_c + FP_c}, \quad R_c = \frac{TP_c}{TP_c + FN_c}. \quad (16)$$

The F-score for each class is the harmonic mean of precision and recall:

$$F_c = \frac{2\,P_c\,R_c}{P_c + R_c}. \quad (17)$$

The weighted F-score was computed as the weighted average of the F-scores for all classes:

$$F_{\text{weighted}} = \sum_{c=1}^{C} w_c\,F_c, \quad (18)$$

where $w_c$ is the proportion of samples in class $c$. This comprehensive evaluation framework ensured that the selected features and filter configurations were assessed on a sound and unbiased basis, providing a reliable guide for future industrial implementation.

## 3. Results and discussion

### 3.1. Analysis of features

Fig. 2 demonstrates the window placement process on mortar samples to extract reflectance data as described by Eq. (2). An RGB image was generated by selecting reflectance values corresponding to the RGB channels, as depicted in Fig. 3. The extracted features, computed according to Eqs. (4), (6), (8), and (9) for the window subset data, are summarized in Table C.1 (Appendix section Appendix C). Due to the differences in scale between the low reflectance values ($F_B$ through $F_{1000}$) and the whole-spectrum features ($F_{\text{peak}}$ and $F_{\text{area}}$), we applied a scaler within the MLP model to normalize the feature values and ensure balanced input for training.

By utilizing automatic masking of the images displayed in the RGB channels through the rembg package, multiple window subsets were identified by placing them on a regular grid to extract features across the entire samples at different locations, as shown in Fig. 3. This figure also presents the mean reflectance curves, which exhibit a consistent pattern across the subsets. This consistency was observed for all materials, as illustrated in Fig. C.9 (Appendix section Appendix C); however, different materials displayed varying degrees of scatter, and differences were also noted between individual samples within the same material class. Understandably, the reflectance curves were low for the asphalt samples, as asphalt is dark gray to black, while the white EPS samples exhibited high reflectivity at all wavelengths. Homogeneous materials such as roof tiles or EPS showed relatively low scatter in their reflectance data. In contrast, materials with variable textures, such as glazed tiles recorded from the top, displayed differences even within a single fragment. This variability was particularly evident in brick fragments containing numerous inhomogeneities, concrete with exposed aggregates of different origins, and wood with a variable texture.

Fig. 4 presents pair plots illustrating the clustering of individual samples when characterized by selected features: (i) $F_B$, representing one of the color channels; (ii) $F_{1000}$, located at the opposite end of the reflectance spectrum; and (iii) $F_{\text{peak}}$ and $F_{\text{area}}$, which represent whole-spectrum features. A notable disadvantage of using $F_{\text{peak}}$ is that spectral peaks often occur at the same wavelengths for different samples, frequently at the beginning or end of the spectrum, which diminishes its discriminatory power. Based on the clustering patterns, the selected features effectively distinguish among highly reflective EPS samples (which appear white in the visible spectrum), wood samples with low reflectance at $\lambda_B$ but relatively high reflectance at $\lambda = 1000$ nm, and dark asphalt specimens exhibiting low reflectance across all recorded wavelengths $\lambda$.

### 3.2. Modeling

All individual MLP models, trained on different sets of features, reached the minimum of the validation loss function $L_{\text{val}}$ after approximately 1 000 iterations, independent of the number of input features, as presented in Fig. D.10, Appendix Appendix D. This convergence behavior indicates that the models required a similar number of iterations to learn the underlying patterns in the data, regardless of the feature set size or composition.

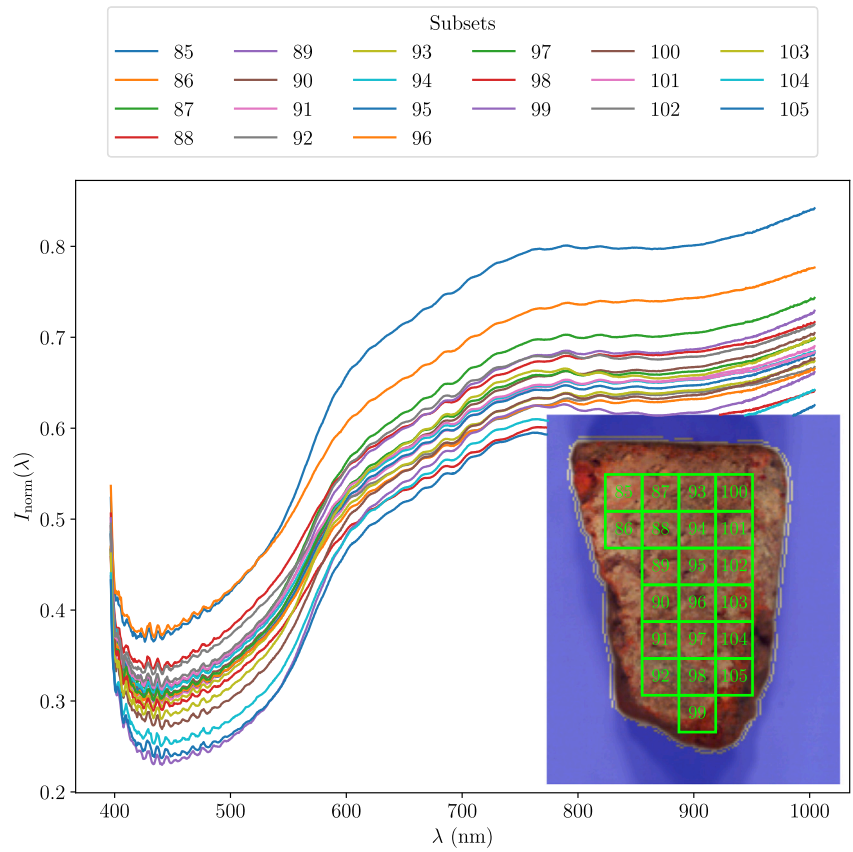

**Fig. 3.** Reflectance curves for individual subsets placed over the brick fragment and illustration of the process of subset placement over a masked ceramic brick fragment; the size of subsets was equal to 20 × 20 px, and the subsets were placed to occupy the entire masked fragment.

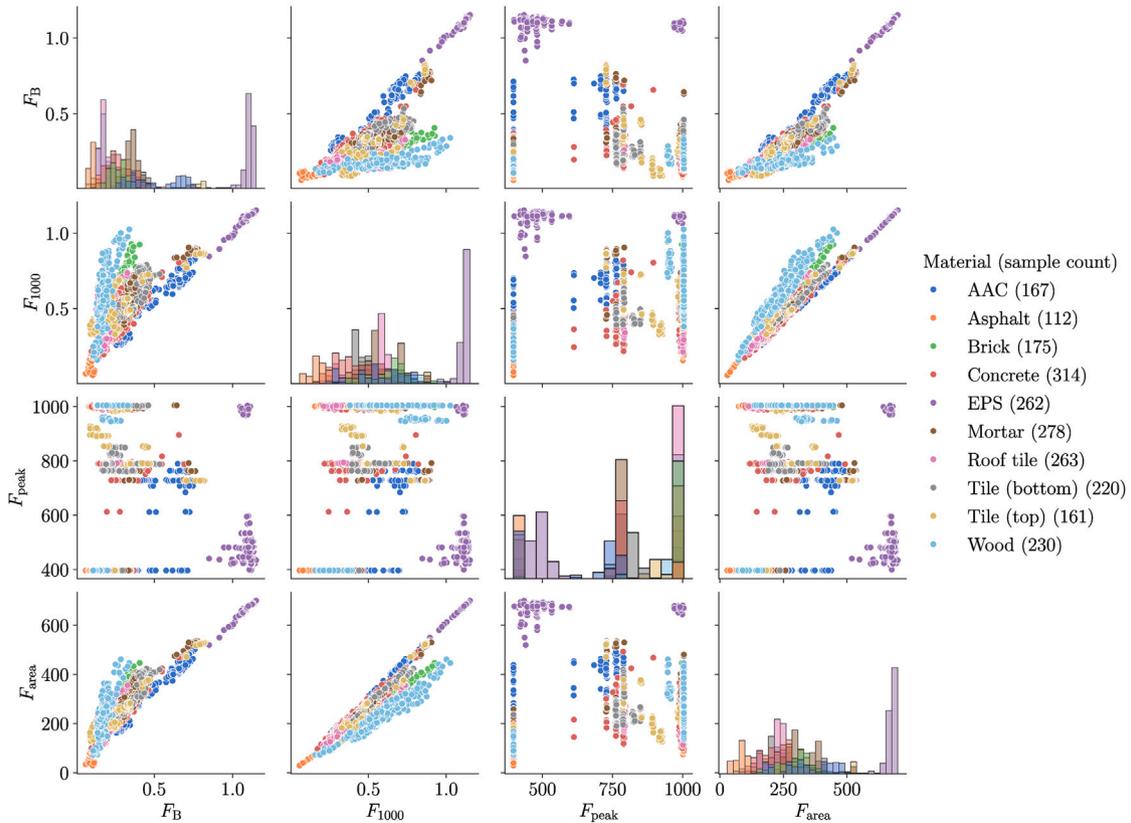

**Fig. 4.** Illustrative pair plot visualizing correlations among selected features used for training the machine learning models; the charts on the diagonal represent the distributions for the respective features.

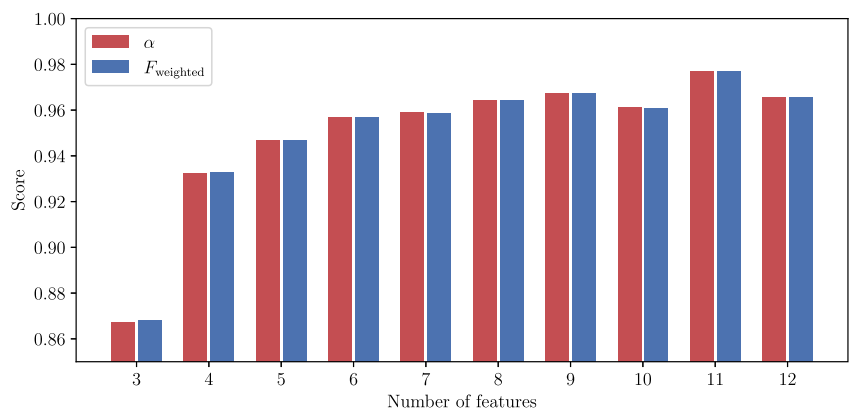

**Fig. 5.** Accuracies and weighted F-scores achieved on the testing dataset using the MLP model; these results were obtained by training the model with different numbers of features, illustrating how the quantity of input features affects performance.

*3.2.1. Number of narrowband filters*

The objective of Stage 1 was to determine the minimum number of wavelengths that need to be recorded to achieve high prediction accuracy and identify which specific wavelengths are most effective. Recording and evaluating the entire spectrum for classifying CDW fragments would be computationally prohibitive and necessitate complex and expensive hardware, rendering it impractical for widespread application. Based on the performance of the MLP models, as illustrated in Fig. 5, it is evident that recording reflectance at more than two additional wavelengths beyond the RGB channels (5 features in total) does not confer significant benefits. Specifically, including more than two supplementary wavelengths did not substantially improve prediction accuracy. Interestingly, the addition of both whole-spectrum features, $F_{peak}$ and $F_{area}$, appeared to confuse the model, resulting in lower prediction accuracies compared to using only $F_{peak}$ or intensities at 9 wavelengths. This finding suggests that simpler models utilizing reflectance at a few strategically chosen wavelengths may be effective and efficient for the classification task.

When trained using only the RGB channels ($F_R$, $F_G$, and $F_B$), the model was able to distinguish between materials with distinct colors; however, grayish materials such as concrete, mortar, and the bottom side of tiles were often confused with one another, as demonstrated by the confusion matrix in Fig. E.12, Appendix Appendix E. Similarly, reddish materials like ceramic brick fragments and roof tiles were frequently misclassified. The model also confused bricks with wood in 11% of cases, indicating limitations in distinguishing materials with similar color profiles using only RGB information.

Adding the normalized intensity $I_{norm}$ at wavelengths $\lambda = 500$ nm and $\lambda = 800$ nm significantly improved the model's accuracy, as shown in Fig. E.12. The lowest classification accuracy was observed for concrete (87%), which was primarily misclassified as the top side of cladding tiles (6%), followed by mortar fragments and the bottom part of tiles. This confusion between concrete, mortar, and the bottom side of cladding tiles is justifiable since all these materials contain cementitious paste, leading to similar reflectance characteristics. Notably, the model was able to distinguish between brick fragments and roof tiles despite their similarity in the RGB channels, highlighting the effectiveness of the additional wavelength features in improving material differentiation.

*3.3. Central wavelengths and FWHM*

Evaluation of classification accuracies for different configurations of $\lambda_{c1}$ and $\lambda_{c2}$ at each considered FWHM value made it possible to identify optimal wavelength combinations and to assess the influence of the bandwidth. When working with commonly available narrowband filters (5–50 nm), the variation in average accuracy $\bar{\alpha}$ across the tested FWHM bandwidths was minimal, ranging from 0.9052 (FWHM = 5 nm) to 0.9068 (FWHM = 35 nm). Notably, the average weighted F-scores $\overline{F}_{weighted}$ closely matched the average accuracies $\bar{\alpha}$. This alignment of accuracy and weighted F-score suggests that the classification performance was balanced across classes, with minimal skew or imbalance affecting the precision–recall relationship.

By smoothing the raw heatmaps (Appendix section Appendix D) using a Gaussian kernel, the underlying trends in wavelength selection became clearer (Fig. 6). These smoothed results indicate that the optimal range for $\lambda_{c1}$ lies between approximately 650 nm and 750 nm, while the optimal range for $\lambda_{c2}$ extends from about 850 nm to 1 000 nm. Similar importance of near-infrared wavelengths for recognizing construction materials has been observed by Ilehag et al. (2017).

Additional insights regarding the effect of different narrowband filter configurations on material discrimination can be obtained from confusion matrices for the evaluated scenarios, as presented in Appendix Appendix E. For instance, these matrices illustrate the performance of a model trained using the RGB features and intensities at $\lambda_{c1} = 400$ nm and $\lambda_{c2} = 624$ nm with a 5 nm FWHM, which yielded a lower accuracy ($\alpha = 0.778$), contrasted against a configuration of $\lambda_{c1} = 775$ nm and $\lambda_{c2} = 975$ nm at a 20 nm FWHM that achieved a higher accuracy ($\alpha = 0.948$). Although this best-performing configuration provided more accurate overall classification, it still misclassified concrete in 19% of cases, often confusing it with the bottom side of cladding tiles (8%). This outcome highlights that even the best filter configurations may not achieve perfect performance for all materials, and that prioritizing accuracy for specific classes may require targeted adjustments to wavelength selection or filter bandwidth.

In summary, while the presented results offer guidance on wavelength and bandwidth selection for improving CDW classification, it is important to consider that these recommendations are tailored to the specific material set analyzed here. If the composition of sorted materials were to change, a similar evaluation would be necessary to ensure that the chosen filter settings remain effective.

**4. Conclusion**

Accurate classification of construction and demolition waste (CDW) is critical for improving recycling efficiency and supporting sustainable management practices. The results presented in this study were derived from a systematic approach aimed at identifying a minimal number of additional wavelengths to supplement standard RGB imaging, and then refining these insights to determine optimal narrowband filter configurations for practical, real-time sorting applications.

Key findings include:

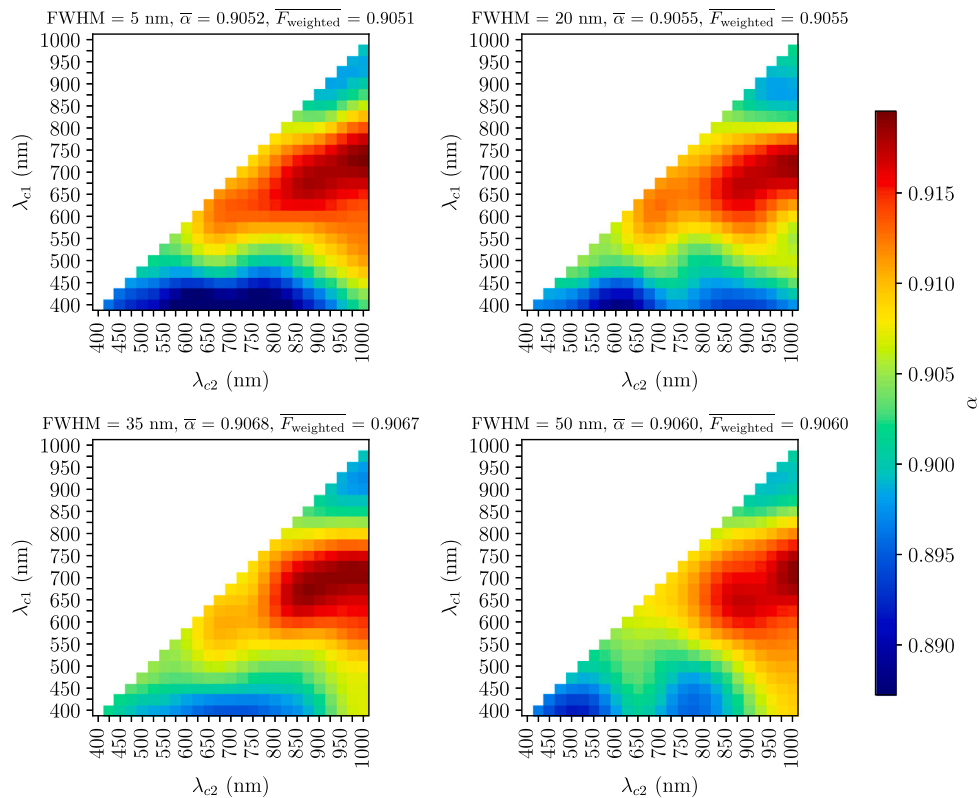

**Fig. 6.** Heatmap of accuracies, $\alpha$, for different combinations of $F_R$, $F_G$, and $F_B$ and two additional features $F_{\lambda_{c1},\text{FWHM}}$ and $F_{\lambda_{c2},\text{FWHM}}$, smoothed using a Gaussian kernel ($\sigma = 2$), indicating the optimal sets of $\lambda_{c1}$ and $\lambda_{c2}$ for different FWHM.

- A minimal set of two additional wavelengths beyond the RGB channels was sufficient to significantly enhance classification accuracy, eliminating the need for full-spectrum hyperspectral data.

- Although certain models could potentially outperform the chosen multilayer perceptron (MLP) classifier, the focus of this research was on selecting suitable wavelengths and filter bandwidths rather than optimizing model complexity. The chosen MLP architecture, determined through preliminary tests, provided a robust and computationally efficient solution.

- The evaluation of narrowband filters at central wavelengths ranging from 425 nm to 975 nm, in increments of 25 nm, and with full-width at half-maximum bandwidths of 5, 20, 35, and 50 nm, showed that classification accuracy remained high and broadly insensitive to the tested bandwidths. This stability indicates flexibility in selecting practical filter configurations from commonly available options.

- Analysis of the accuracy heatmaps revealed an optimal range of approximately 650–750 nm for the first additional wavelength and 850–1000 nm for the second. This outcome emphasizes the importance of near-infrared regions for material discrimination.

- Although the highest achieved accuracies were favorable, certain CDW material classes, such as concrete, remained challenging. Confusion matrices demonstrated residual misclassifications that indicate the need for further refinement if the classification priorities shift toward different materials or performance metrics.

This study provides a novel, two-stage methodology that bridges the gap between comprehensive but impractical hyperspectral imaging and cost-effective multispectral solutions for large-scale, real-time sorting of construction and demolition waste. The recommended wavelengths and filter parameters serve as actionable guidelines, but it is important to recognize that any significant change in material composition would necessitate repeating the analysis for selecting optimum narrowband filters.

**CRediT authorship contribution statement**

**Stanislav Vítek:** Validation, Supervision, Investigation, Formal analysis, Data curation, Conceptualization. **Tomáš Zbíral:** Investigation, Data curation. **Václav Nežerka:** Writing – original draft, Visualization, Software, Methodology, Investigation, Conceptualization.

**Funding**

This work was funded by the European Union under the project ROBOPROX (no. CZ.02.01.01/00/22_008/0004590), by the European Union's Horizon Europe Framework Programme (call HORIZON-CL4-2021-TWIN-TRANSITION-01-11) under grant agreement No. 10105 8580, project RECONMATIC (Automated solutions for sustainable and circular construction and demolition waste management), and the Czech Technical University in Prague, grant agreement No. SGS24/003/OHK1/1T/11 (Application of state-of-the-art technologies for enhancing sustainability of the construction sector).

**Declaration of competing interest**

All the authors of the submitted manuscript titled "Identification of critical wavelengths for classifying construction and demolition waste materials using hyperspectral imaging" certify that they have NO affiliations with or involvement in any organization or entity with any financial interest (such as honoraria; educational grants; participation in speakers' bureaus; membership, employment, consultancies, stock ownership, or other equity interest; and expert testimony or patent-licensing arrangements), or non-financial interest (such as personal or professional relationships, affiliations, knowledge or beliefs) in the subject matter or materials discussed in this manuscript.

**Appendix A. Samples**

See Fig. A.7.

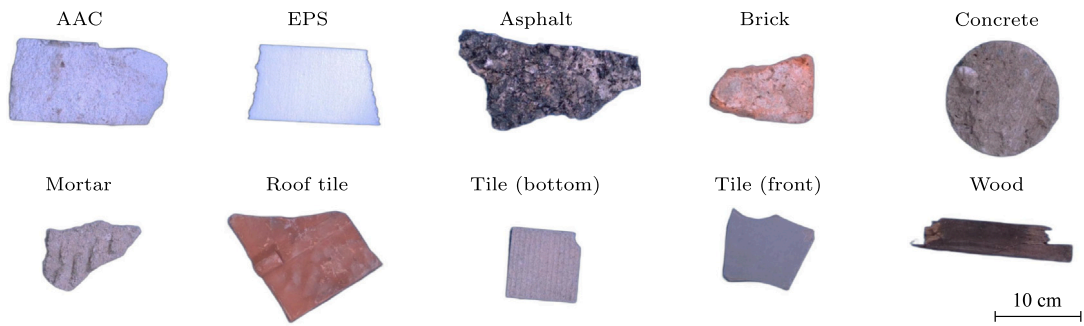

**Fig. A.7.** Representative samples of CDW materials used in this study.

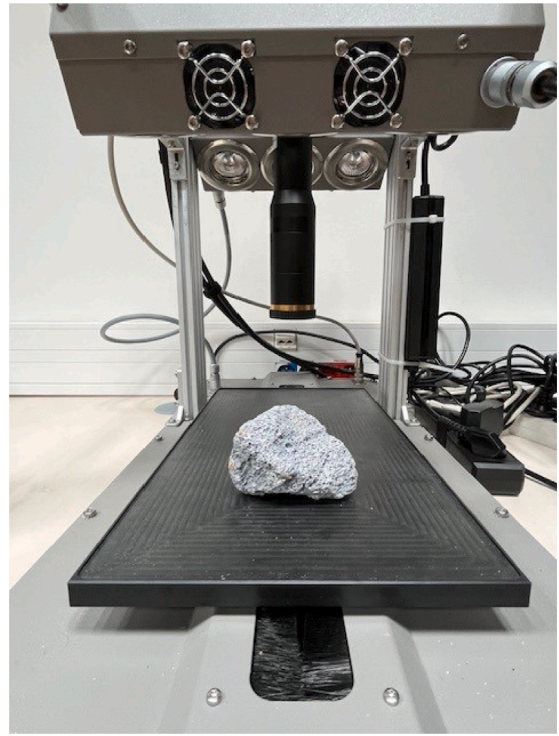

**Fig. B.8.** Data acquisition using the hyperspectral camera at the Faculty of Electrical Engineering, CTU in Prague.

**Table C.1**
Features extracted from the reflectance curves provided in Fig. 2.

|  | mortar-01 | mortar-02 | mortar-03 | mortar-04 |
| --- | --- | --- | --- | --- |
| $F_B$ | 0.3662 | 0.3731 | 0.3516 | 0.4091 |
| $F_G$ | 0.545 | 0.5376 | 0.4694 | 0.484 |
| $F_R$ | 0.6644 | 0.6527 | 0.5398 | 0.5278 |
| $F_{400}$ | 0.41 | 0.4011 | 0.3763 | 0.4198 |
| $F_{500}$ | 0.4642 | 0.4623 | 0.4181 | 0.4564 |
| $F_{600}$ | 0.6407 | 0.6266 | 0.5234 | 0.5143 |
| $F_{700}$ | 0.6757 | 0.6658 | 0.5466 | 0.5325 |
| $F_{800}$ | 0.6905 | 0.6871 | 0.5547 | 0.5392 |
| $F_{900}$ | 0.6842 | 0.687 | 0.5496 | 0.5438 |
| $F_{1000}$ | 0.6884 | 0.695 | 0.5494 | 0.5528 |
| $F_{peak}$ | 789.31 | 1004.2 | 789.31 | 993.03 |
| $F_{area}$ | 373.695 | 371.46 | 308.637 | 310.7474 |

**Appendix B. Equipment**

See Fig. B.8.

**Appendix C. Reflectance data**

See Fig. C.9 and Table C.1.

**Appendix D. Model training**

See Figs. D.10 and D.11 and Table D.2.

**Appendix E. Confusion matrices**

See Figs. E.12 and E.13.

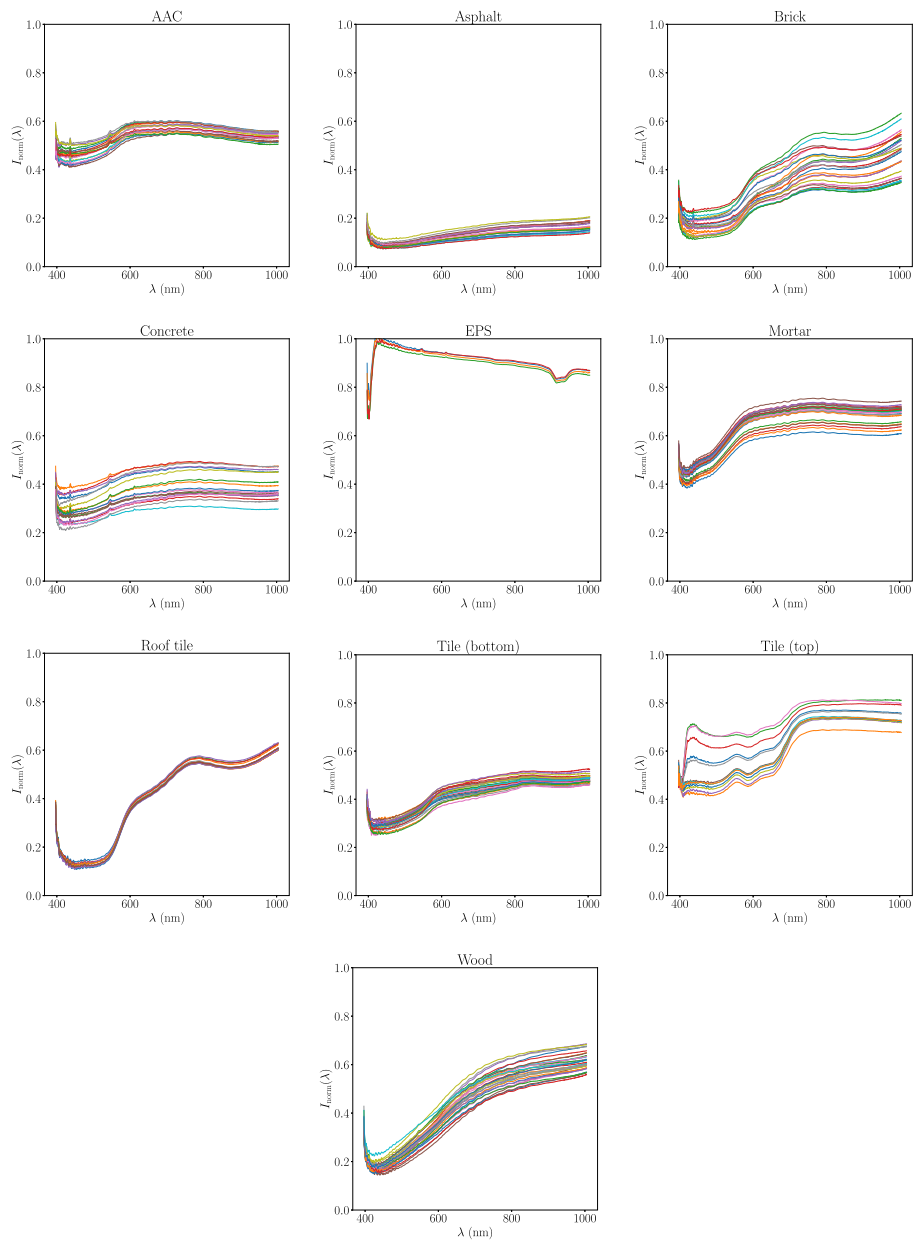

**Fig. C.9.** Illustration of the differences in the reflectance curves for individual materials, demonstrated on selected samples.

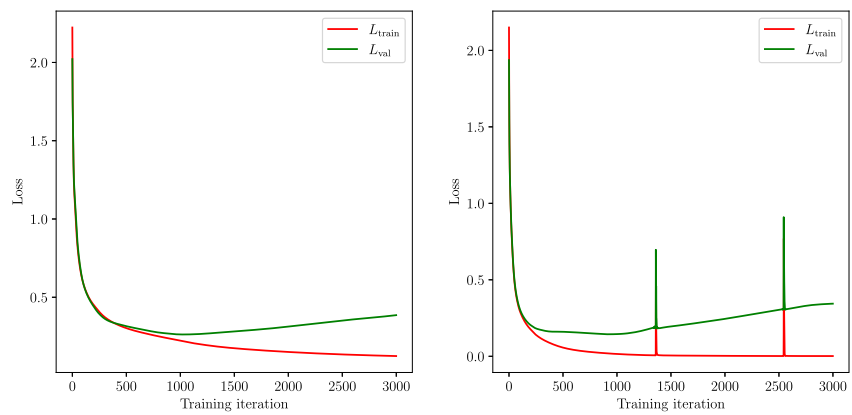

**Fig. D.10.** Learning curves for the model trained using only the RGB channels, $F_R$, $F_G$, and $F_B$ (left), and the model trained using all the features available, including $F_{peak}$ and $F_{area}$ (right).

Table D.2
Summary of input parameters for the MLP classifier implemented in Scikit-Learn v1.1.3 (neural_network.MLPClassifier class).

| Input parameter | Keyword argument | Value | Description |
| --- | --- | --- | --- |
| Random state | random_state | 0 | Ensures deterministic behavior during fitting |
| Learning rate | learning_rate_init | 0.015 | Controls the step size in updating weights |
| Maximum iterations | max_iter | 800 | Maximum number of training epochs |
| Learning rate schedule | learning_rate | 'constant' | Uses a constant learning rate |
| Solver | solver | 'adam' | Optimization algorithm (Kingma and Ba, 2015) |
| Activation function | activation | 'tanh' | Activation function for the hidden layer |
| Hidden layer sizes | hidden_layer_sizes | (20, 20) | Two hidden layer with 20 neurons |

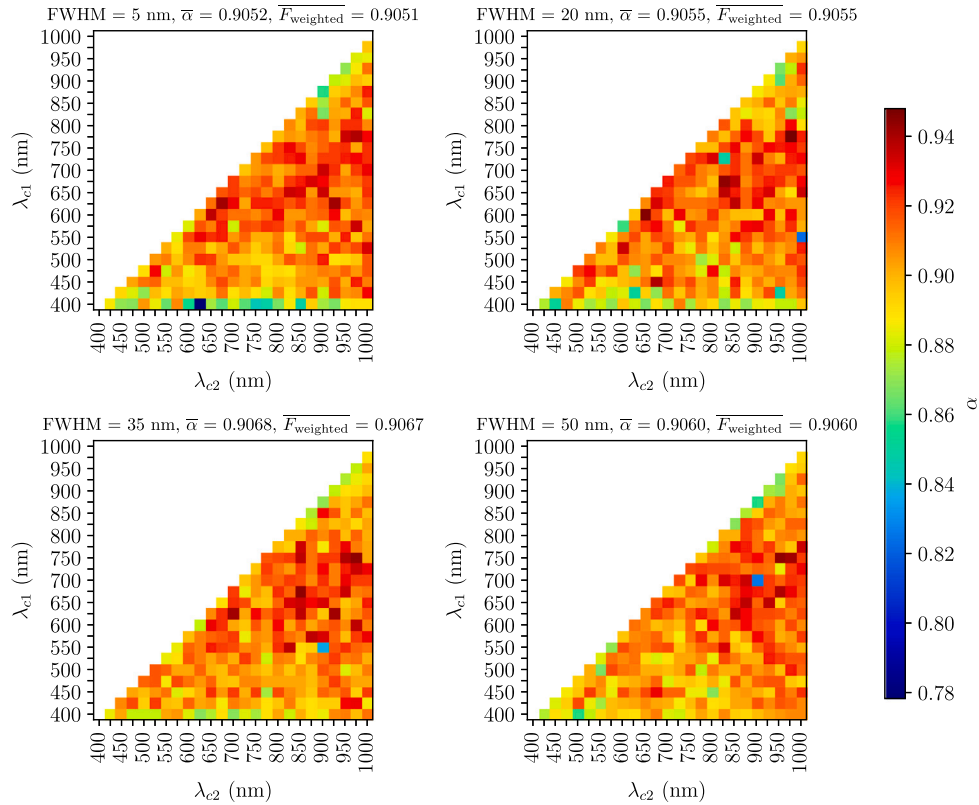

**Fig. D.11.** Heatmap of accuracies, $\alpha$, for different combinations of $F_R$, $F_G$, and $F_B$ and two additional features $F_{\lambda_{c1},\text{FWHM}}$ and $F_{\lambda_{c2},\text{FWHM}}$, indicating the optimal subset of $\lambda_{c1}$ and $\lambda_{c2}$ for different FWHM.

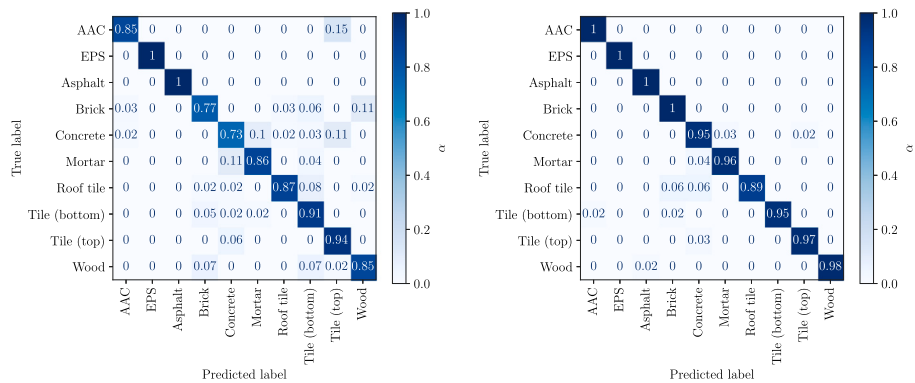

**Fig. E.12.** Confusion matrices showing the accuracy for the testing dataset evaluated using the model trained on the RGB channels ($F_R$, $F_G$, and $F_B$) (left) and trained using all the features available, including $F_{\text{peak}}$ and $F_{\text{area}}$ (right).

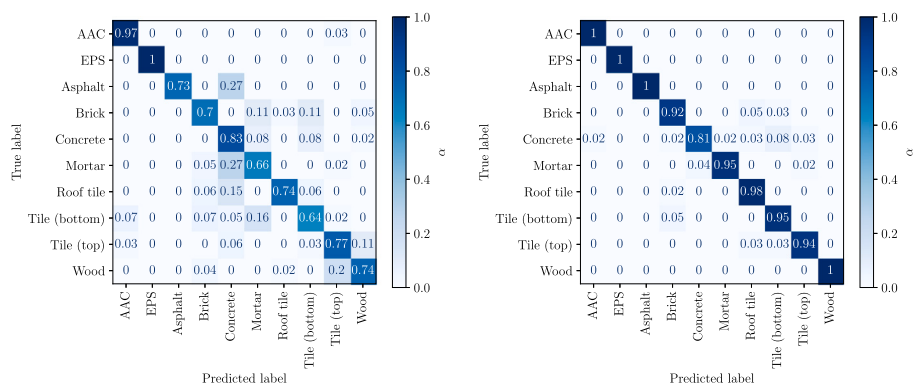

**Fig. E.13.** Confusion matrices showing the accuracy for the testing dataset evaluated using the model trained on the RGB channels ($F_R$, $F_G$, and $F_B$) and intensities for (i) $\lambda_{c1}$ = 400 nm, $\lambda_{c2}$ = 624 nm, and FWHM = 5 nm (the worst combination yielding $\alpha$ = 0.778, left) and (ii) $\lambda_{c1}$ = 775 nm, $\lambda_{c2}$ = 975 nm, and FWHM = 20 nm (the best combination yielding $\alpha$ = 0.948, right).

**Data availability**

The link to the data and scripts is shared within the paper.